# High-efficiency thermophotovoltaic system that employs an emitter based on a silicon rod-type photonic crystal

Masahiro Suemitsu[1,2], Takashi Asano[1*], Takuya Inoue[1], Susumu Noda[1]

[1]Department of Electronic Science and Engineering, Kyoto University, Kyoto 606-8501, Japan

[2]Energy Technology Laboratories, Osaka Gas Co., Ltd., 6-19-9 Torishima, Konohana-Ku 554-0051, Osaka, Japan

*Corresponding author: tasano@qoe.kuee.kyoto-u.ac.jp

**Abstract**

Thermophotovoltaic systems in principle enable utilization of heat that is usually regarded as wasted energy. However, the wavelength selectivity of the thermal emitter required for high efficiencies is rather difficult to control with conventional designs. Here, we design a thermophotovoltaic system comprising silicon rods as thermal emitter with a relatively narrow emission spectrum and a photovoltaic cell with a band gap corresponding to 1.76 μm, and verify efficient power generation. By accurately measuring the heat flux that enters the emitter, the emitter temperature, and the electrical output power of the photovoltaic cell, we find that the actual system efficiency (ratio of ingoing heat flux to output power) is 11.2% at an emitter temperature of 1338 K, and that the output power density footprint is 0.368 W/cm$^2$. The obtained efficiency is relatively high, i.e., 1.65 times that of the previously reported record value (6.8%). Further efficiency improvements in the future may lead to development of distributed energy supplies using combustion heat.

**Introduction**

   Control of the thermal emission spectrum from a hot object has been considered for applications in various fields (*1–13*). It is particularly important for applications such as thermophotovoltaics (*14,15*). A thermophotovoltaic system allows us to generate electricity from thermal energy by converting thermal radiation emitted from a hot object (the so-called thermal emitter) into electricity by a photovoltaic (PV) cell. Because such systems have the potential to provide conversion efficiencies and output power densities per volume comparable to that of fuel cells if the thermal emitter is designed correctly, they have been attracting attention in recent years (*15–21*). The most fundamental parameter for the performance of a thermophotovoltaic system is the relation between the emission spectrum of the thermal emitter and the band-gap energy of the absorber in the PV cell. To illustrate the importance of this relation, first, consider the case where the input heat flux with power density $P_{in}$ is converted into black-body radiation without losses, and this radiation is converted into electricity by an ideal single-junction PV cell with a band-gap energy defined by the corresponding wavelength $\lambda_{Eg}$ (Fig. 1A). The PV cell is placed adjacent to the emitter, and they have the same size. In this case, the light with wavelengths longer than $\lambda_{Eg}$ (the region L) is not absorbed by the PV cell, and only the short-wavelength light (in region S) is absorbed. Hence, the ratio of absorbed energy to the total radiated energy increases as the wavelength $\lambda_{Eg}$ becomes longer. This ratio is the so-called spectral efficiency denoted by $\eta_{spectrum}$ and shown with the red curve in Fig. 1B. For instance, the red curve in Fig. 1B shows that in case of a black-body with a temperature of 1300 K, the spectral efficiency is limited to about 25% for $\lambda_{Eg}$ near the peak wavelength of the black-body radiation (≈ 2 μm), but it increases to about 80% for $\lambda_{Eg}$ = 5 μm. However, the losses through hot-carrier relaxation to the band edge in the PV cell's absorber increase as the wavelength of the absorbed light becomes shorter. Therefore, the internal conversion efficiency $\eta_{int}$ (i.e., the fraction of absorbed power that is actually converted into electrical power with power density $P_{out}$) decreases for larger $\lambda_{Eg}$ as shown with the blue curve in Fig. 1B. Because the system efficiency defined by $\eta_{system} = P_{out}/P_{in}$ can be also written as the product of spectral and internal efficiencies, the maximum efficiency of about 30% is obtained at $\lambda_{Eg} \approx$ 4 to 6 μm (Fig. 1B; black broken curve). Here, a large $P_{out}$ (green curve) of more than 4 W/cm² can be obtained. A perfect spectral efficiency would be obtained if the emission spectrum has a shape like that shown with the red curve in Fig. 1C; if a wavelength-selective emitter that perfectly cuts the long-wavelength contribution above $\lambda_{Eg}$ is available, $\eta_{spectrum}$ can reach 100%. In this case, the $\eta_{system}$ is solely determined by $\eta_{int}$ as shown in Fig. 1D.

Furthermore, the $P_{in}$ required to maintain a certain emitter temperature would be reduced by a factor of S/(S+L). Then we can expect a high system efficiency of more than 57% at $\lambda_{Eg}$ = 1.8 μm and a large $P_{out}$ on the order of 1 W/cm$^2$. However, while this concept is attractive from the viewpoint of efficiency, the actual control of the thermal emission spectrum is difficult. Furthermore, since there exist losses such as heat conduction from the emitter to other device parts, and also factors like the non-ideal properties of the PV cell, the actual conversion efficiencies of real thermophotovoltaic systems ($\eta_{system\_exp}$) are still at a low level.

Until about 2013, the reported experimental system efficiencies, in which the input power has been directly measured, were around 1% (*16–18*, *22–24*). Then, beginning with 2015, $\eta_{system\_exp}$ exceeding 5% started to appear in the literature (*19–21*). For example, Kohiyama et al. have reported 5.1% for $\eta_{system\_exp}$ by employing a ≈1650 K hot Mo-MIM structure for the emitter and a GaSb PV cell (*25*) with a $\lambda_{Eg}$ of 1.85 μm (*21*). Biermann et al. have reported $\eta_{system\_exp}$ = 6.8% by employing a ≈1300 K hot Si-MIM structure on a tungsten substrate for the emitter (*26*). and a InGaAsSb PV cell (*27*) with a $\lambda_{Eg}$ of 2.2 μm (*20*). The important point is that, both publications have estimated losses such as those accompanying heat conduction, which can be reduced by upscaling, and have presented a system efficiency of about 10% by removing these losses from the denominator in $\eta_{system\_exp}$. Novel emitter designs may enable even higher efficiencies. Semiconductor based photonic crystal (PC) structures seem to be advantageous for thermal emitters, because anomalous photonic density of states of a PC and an absorption spectrum of a semiconductor can be utilized to tune the emission spectrum.

In this work, we designed a thermophotovoltaic system that employs a Si rod-type PC, whose emission profile can be controlled better than those of the metallic thermal emitters in the previous reports. Owing to the photonic density of states of the rod-type PC and the step-like absorption spectrum of Si (*11*), we obtain a highly suitable emission spectrum for the employed InGaAs single-junction device. With this we experimentally obtain $\eta_{system\_exp}$ = 11.2% under an emitter temperature of 1338 K. For the system efficiency excluding the losses that can be reduced via upscaling we obtain 14.5%, and the efficiency that can be expected by further improvements of the contact grid design of the PV cell is 19.4%. With respect to thermophotovoltaic systems in which the input power has been directly measured, our experimentally determined efficiency can be considered as the present world record. Furthermore, the predicted upscaled system efficiency exceeds the previously reported semi-empirical limit.

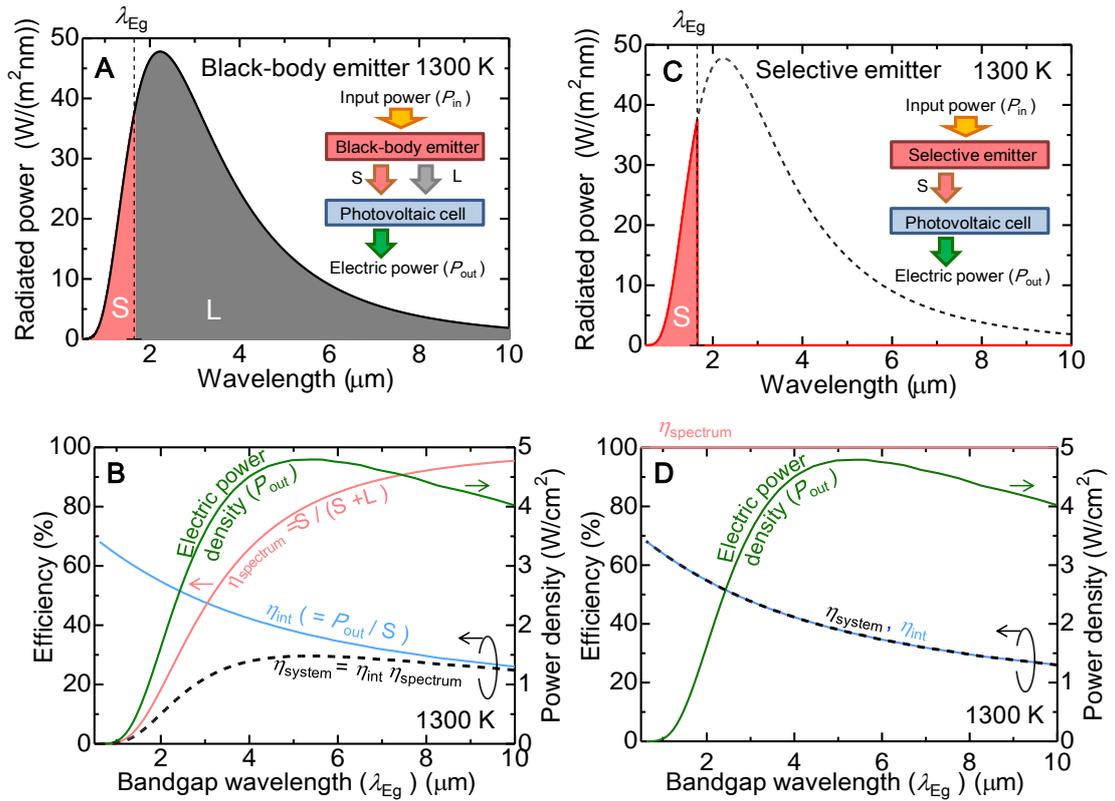

**Fig. 1. Differences between a black-body emitter and a selective emitter.** (**A**) Spectral power density of a black-body emitter. The inset shows a schematic for the case of lossless conversion of the input power into black-body radiation, and the conversion of this radiation into electrical power by an ideal single-junction PV cell. (**B**) The conversion efficiency and output power density as a function of the wavelength equivalent to the band-gap energy of the PV cell in case of the thermophotovoltaic conversion shown in the inset of (**A**). (**C**) Spectral power density of an emitter whose spectrum is optimized for conversion. The inset shows a schematic for the case of the conversion of radiation whose wavelength contribution above $\lambda_{Eg}$ has been cut, and the conversion of this radiation into electrical power by an ideal single-junction PV cell. (**D**) The conversion efficiency and output power density as a function of the wavelength equivalent to the band-gap energy of the PV cell in case of the thermophotovoltaic conversion shown in the inset of (**C**).

**Results**

The thermal emitter used in the present thermophotovoltaic system is a rod-type PC with a rectangular lattice (*11*, *13*) As shown in the illustration in Fig. 2A, this PC was fabricated by processing a polycrystalline Si thin film on a MgO substrate, which is transparent for infrared light. We prepared a 20-nm-thick $HfO_2$ layer on the 500-μm-thick MgO substrate and then grew 825 nm of polycrystalline Si using the LP-CVD method. The PC pattern was written by electron-beam lithography, and the Si rod structure was obtained by dry etching. The PC's geometry is defined by $d = 360$ nm, $h = 825$ nm, and $a = 700$ nm as shown in Fig. 2B. The thermal radiation spectrum that is observed along the surface normal direction from the thermal emitter with this structure when it is heated up to 1300 K, is shown with the red solid curve in Fig. 2C. Additionally, the blue curve in Fig. 2C is the theoretically determined thermal radiation spectrum along the surface normal direction for the ideal rod structure in Fig. 2A. While there are some deviations caused by the difference between the ideal structure and the fabricated structure, both spectra agree fairly well. Because the actual radiation that is incident on the PV cell is the hemispherically integrated radiation, we show its theoretical value in Fig. 2D (since the experimental integration over $2\pi$ sr is difficult). The peak wavelength is about 1.6 μm, and we find that the contribution with wavelengths longer than 1.76 μm is strongly suppressed when compared with the black-body radiation.

For the PV cell we employ a single-junction device based on InGaAs. The photograph of the fabricated PV cell is shown in Fig. 3A, and the measured external quantum efficiency is shown in Fig. 3B. The light receiving surface of the PV cell is about 36 $mm^2$, the cell's aperture ratio (ratio of active area to total light receiving area) is 77%, the series resistance is 0.01 $\Omega cm^2$, the shunt resistance is 1000 $\Omega cm^2$, the dark current is $5\times10^{-6}$ $A/cm^2$, and the diode's ideality factor $n$ is 1.5. As can be seen in Fig. 3B, the $\lambda_{Eg}$ of the absorber layer is about 1.76 μm, and thus the $\eta_{spectrum}$ with respect to the spectrum in Fig. 2C is 42.6%. For comparison we note that, since the $\eta_{spectrum}$ of this PV cell under direct illumination with black-body radiation at 1300 K is 11.5%, $\eta_{spectrum}$ increased by a factor of more than three due to the control of the thermal emission via the Si rod structure.

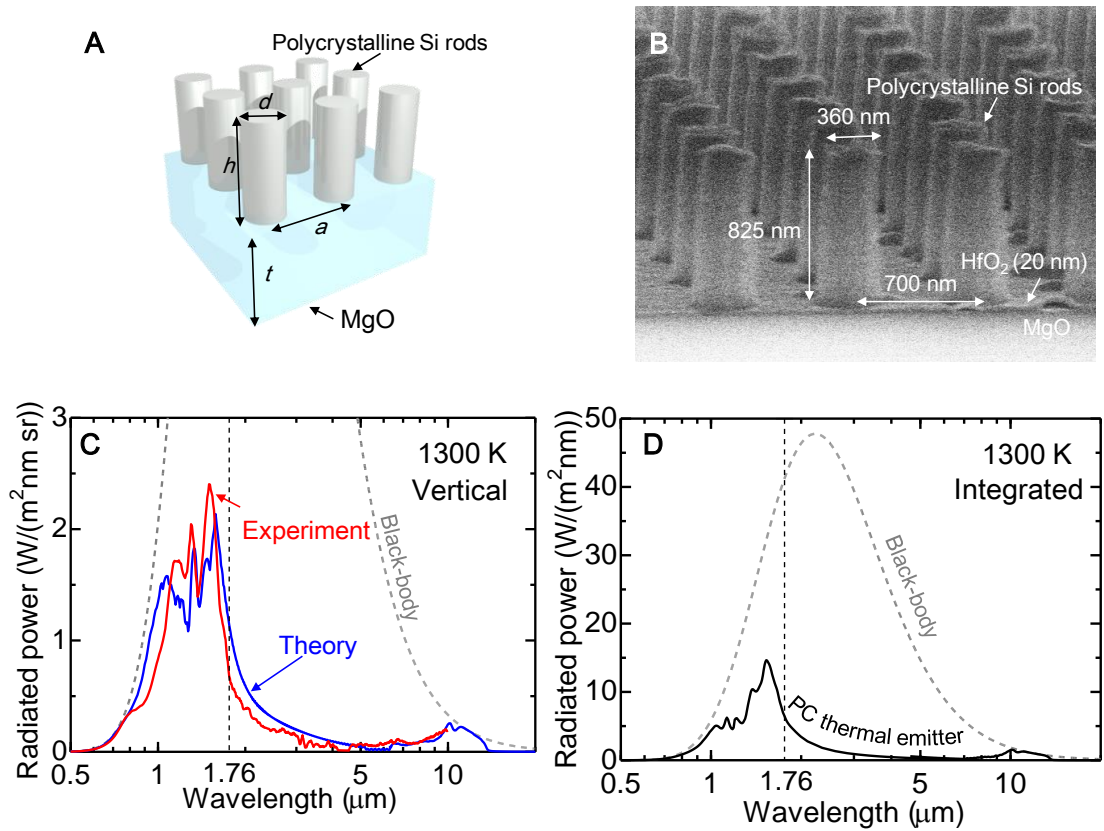

**Fig. 2. Properties of the PC thermal emitter.** (**A**) Illustration of the PC thermal emitter. (**B**) Scanning electron microscope image of the Si rod-type PC before atomic layer deposition. (**C**) The experimental (red curve) and theoretical (blue curve) thermal radiation spectrum along the surface normal direction of the emitter. (**D**) The hemispherically integrated spectrum of the theoretical thermal radiation of the thermal emitter at 1300 K (black solid curve), and the hemispherically integrated black-body radiation spectrum for 1300 K (gray broken curve).

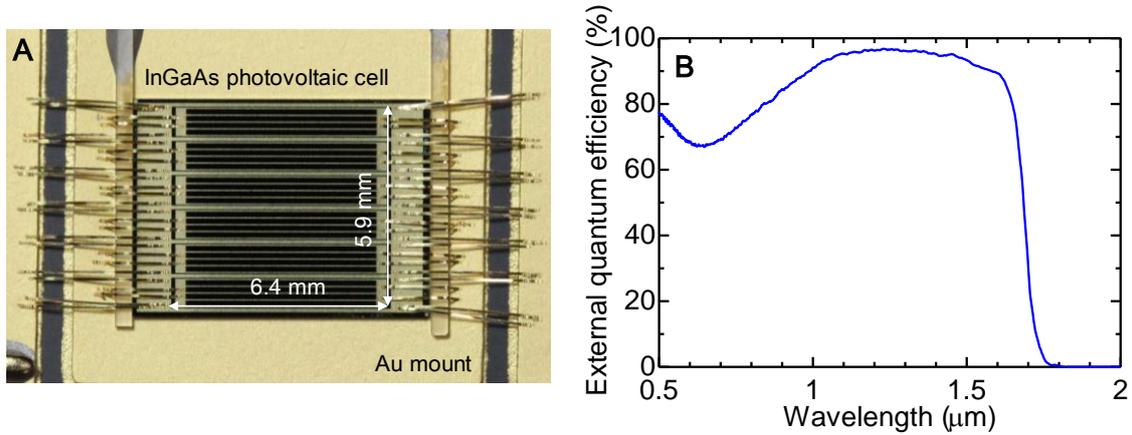

**Fig. 3. PV cell for electricity generation.** (**A**) The visible structure of the fabricated PV cell. The thin lines on the cell is the front surface contact grid, which is connected to the large contact pads via several gold ribbons. (**B**) External quantum efficiency of the PV cell.

By using the above thermal emitter and PV cell, we constructed a thermophotovoltaic system that is optimized for accurate characterization as schematically shown in Fig. 4A. For actual implementation we consider a layout as shown in Fig. 4B; combustion heat or solar heat is transferred to the emitter surface via conduction from the edge, and PV cells are placed on both sides of the emitter. In the present system for characterization, an internal heater (Ohmic resistor controlled via 4 electrical contacts) supplies the emitter with heat, and the thermal radiation from the emitter is converted into electrical power by the PV cells. The present setup also enables direct measurement of the emitter temperature by forming a micro thermocouple on the emitter. As the Si rod-type PC structure is situated on top of a transparent MgO substrate, almost the same thermal radiation spectrum is obtained on both sides of the emitter. Therefore, PV cells are placed on both sides of the emitter. In efficiency measurements of thermophotovoltaic systems, the quantitative evaluation of the total input power ($P_{\text{in}}^{\text{tot}}$) and the total power that is not emitted towards the PV cells ($P_{\text{leak}}^{\text{tot}}$) is difficult in general. The present setup enables a quantitative evaluation of $P_{\text{in}}^{\text{tot}}$ and $P_{\text{leak}}^{\text{tot}}$ by suspending the emitter in vacuum with thin Pt wires (for thermal insulation) and converting the electrical energy transmitted through the thin wires to thermal energy via the internal heater. The $P_{\text{in}}^{\text{tot}}$, the total output power from the two PV cells $P_{\text{out}}^{\text{tot}}$, and the emitter temperature can be measured at the same time, which is considered essential for reliable results. The photograph of the thermal emitter including the heater is shown in Fig. 4C. The Si rod-type PC is formed on a 4.8×4.8-mm large MgO substrate with a thickness of 65 μm, and the Pt/Ti heater is M-shaped with a total length of 14.3 mm, a width of 10 μm, and a thickness of 300 nm. We note that the ratio of the heater's area to the PC area is 0.6%, and thus the directly generated emission of the latter can be neglected when compared to the emission of the PC. Furthermore, for an accurate temperature measurement, a 200-μm squared Pt/Ti pad was prepared on the side opposite to the contacts for the heater, and connected with 25-mm long Pt and $Pt_{0.9}Rh_{0.1}$ wires with 25 μm in diameter to directly fabricate an S-type thermocouple on the emitter. In the following, the power generation is tested after placing PV cells on both sides of the emitter and reducing their distance to less than 500 μm as shown in Fig. 4D. The view factor between the thermal emitter and the PV cells in this configuration is about 0.97. The PV cells are mounted on copper heat sinks for cooling, and the back-surface temperature is adjusted to about 283 K. During the measurements, the whole system is placed in a vacuum chamber (about $3\times10^{-2}$ Pa).

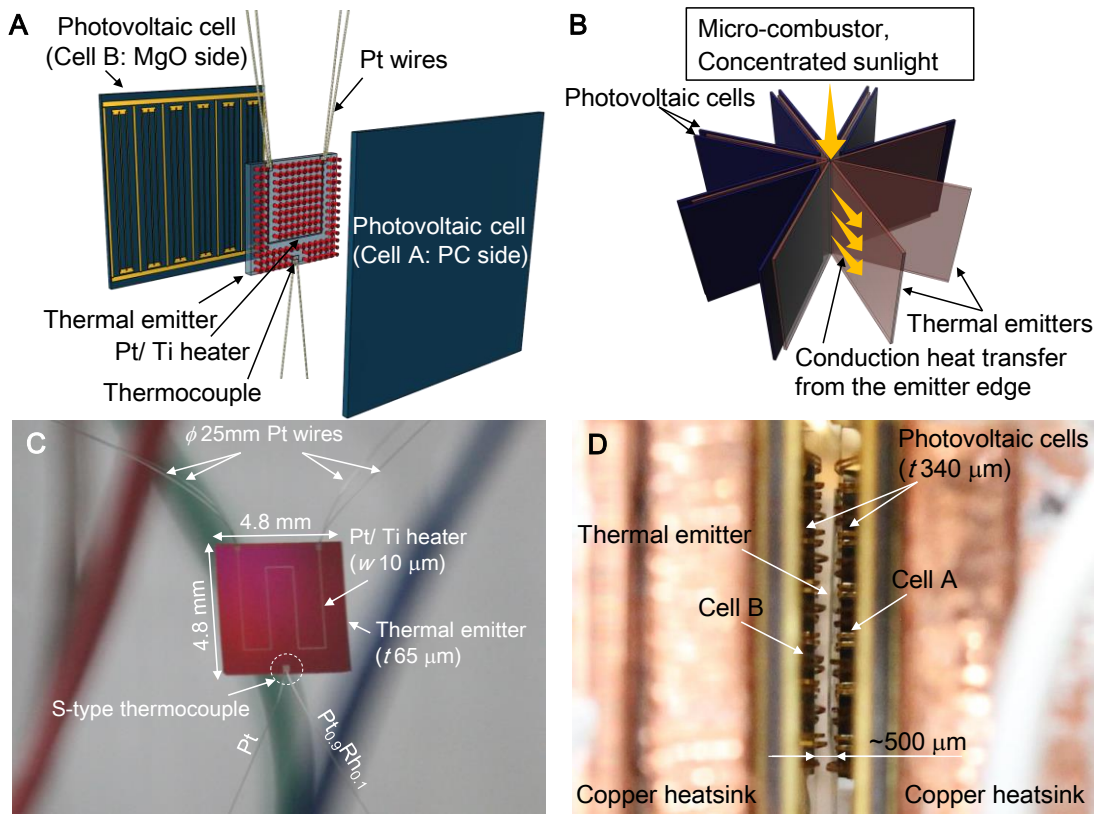

**Fig. 4. Thermophotovoltaic system that employs an emitter based on a Si rod-type PC.** (**A**) Illustration of the designed thermophotovoltaic system. (**B**) The proposed layout for actual implementation. (**C**) Photograph of the fabricated thermal emitter. The heating element and the temperature measurement device are also visible. (**D**) The side view of the fabricated thermophotovoltaic device.

In the present experiment we applied electrical power to the heater in the emitter and let the emitter temperature increase stepwise while measuring the output characteristics of the PC cells at each temperature. The relation between the $P_{in}^{tot}$ applied to the emitter and the emitter temperature is plotted with the blue circles in Fig. 5A. The emitter temperature reaches 1338 K for $P_{in}^{tot}$ = 1.24 W, which is the maximum applied power in this experiment. The inset of Fig. 5A presents the side view of the system at 1338 K. The current–voltage (I–V) characteristics of the PV cells for an emitter temperature of 1338 K are presented in Fig. 5B. We find that almost the same I–V characteristics are obtained for both Cell A (on the PC side) and Cell B (on the MgO substrate side). The total output power from both devices at the optimum operating point is $P_{out}^{tot}$ = 0.139 W. The $P_{out}^{tot}$ as a function of the emitter temperature is shown in Fig. 5A with the black triangles. The trend for the system efficiency $\eta_{system\_exp}$, which is obtained by dividing $P_{out}^{tot}$ by $P_{in}^{tot}$, is plotted with the red circles in Fig 5C. The increase of the emitter temperature is accompanied by an increase in $\eta_{system\_exp}$, and we obtain an efficiency of 11.2% at 1338 K. This experimentally obtained efficiency is 1.65 times larger than the maximum value among the $\eta_{system\_exp}$ that have been previously reported (6.8%) (*15, 20*).

The $P_{out}$ of the system, which is defined as $P_{out}^{tot}$ divided by the total area of the two PV cells, is plotted with black triangles in Fig. 5C. The $P_{out}$ is about 0.184 W/cm$^2$ at 1338 K, but the output power density footprint $P_{out}^{footptint}$ doubles (0.368 W/cm$^2$) because the system has two PV cells. Note that the size of each PV cell ($A_{PV\ cell}$ = 0.378 cm$^2$) is larger than that of the emitter ($A_{emitter}$ = 0.230 cm$^2$) in this prototype system to maintain a large view factor. When we scale up the system, we can make $A_{PV\ cell}$ the same as $A_{emitter}$ without degrading the view factor. $P_{out}^{footptint}$ for this case is mentioned in the discussion section.

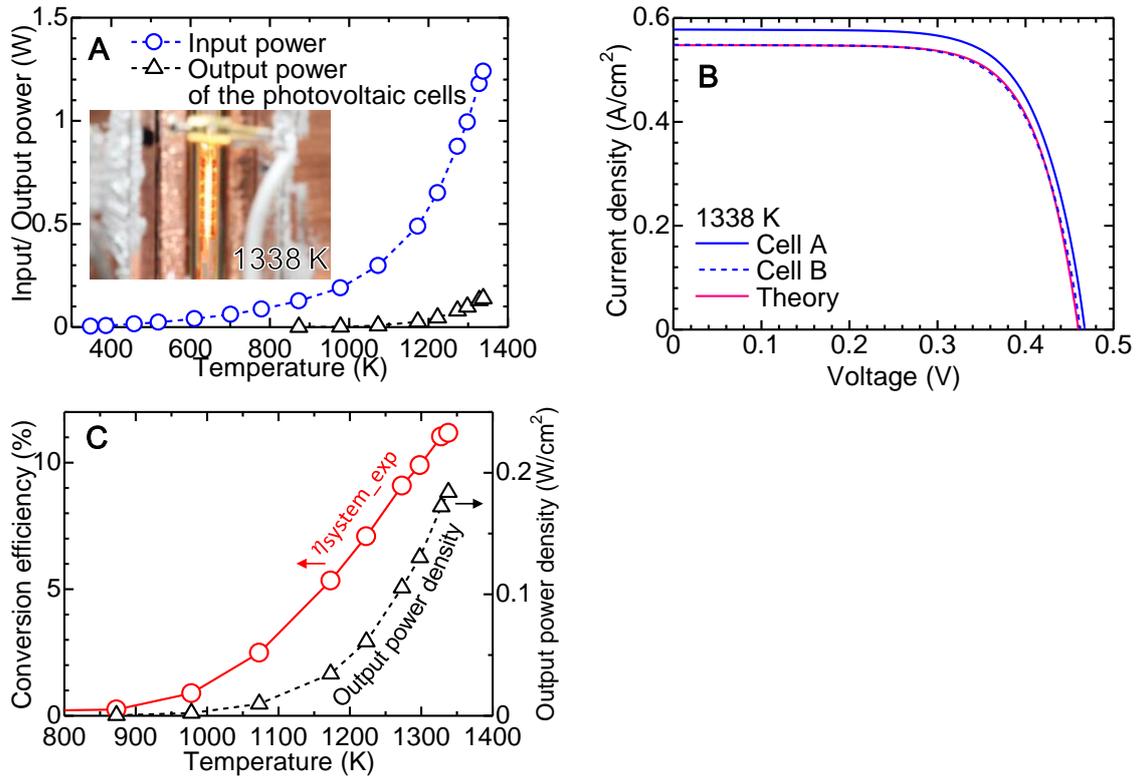

**Fig. 5. Performance of the present thermophotovoltaic system.** (**A**) The heater's input power ($P_{in}^{tot}$ blue circles) and the maximum total output power provided by the PV cells ($P_{out}^{tot}$, black triangles). The inset shows the side view of the fabricated thermophotovoltaic device under operation (emitter at 1338 K). (**B**) The theoretically and experimentally determined I–V characteristics of Cell A and Cell B for an emitter temperature of 1338 K are shown with the red and blue curves, respectively. (**C**) The dependences of $\eta_{system\_exp}$ (red circles) and the output power density (black triangles) on the emitter temperature.

**Discussion**

The energy balance of the present power generation system for an emitter temperature of 1338 K is shown in Fig. 6. Losses such as the radiation from the edges of the thermal emitter, the losses via thermal radiation from the thin Pt wires which are heated up by the emitter, as well as their heat conduction, and also losses caused by the form factor exist. Because these losses can be reduced by increasing the emitter and PV cell areas, we quantitatively evaluate these energy losses ($P_{\text{leak}}^{\text{tot}}$) and discuss the system efficiency that can be achieved after a scale up of the present power generation system by excluding $P_{\text{leak}}^{\text{tot}}$ from $P_{\text{in}}^{\text{tot}}$. Also, when we scale up the system, we can make $A_{\text{PV cell}}$ the same as $A_{\text{emitter}}$ without degrading the view factor.

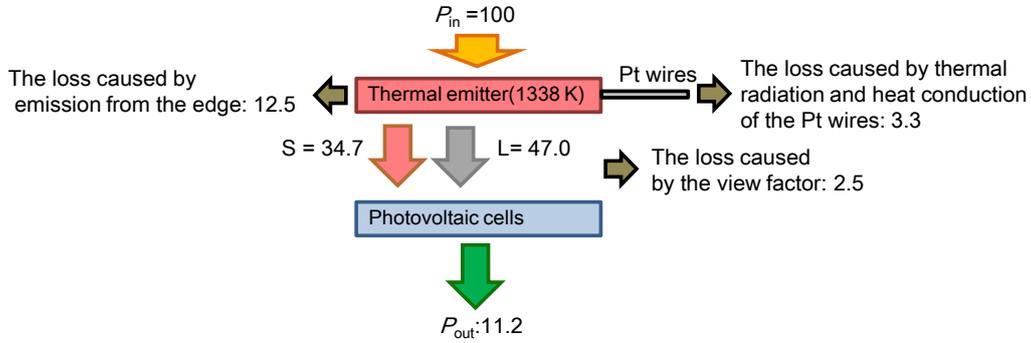

**Fig. 6. The energy balance of our thermophotovoltaic system for an emitter temperature of 1338 K**.

Obviously, the thermal radiation that exits the emitter from its edges cannot be used as it does not strike the PV cell. This power can be determined from the total area of the edges and their emissivity as well as the emitter temperature. Firstly, the area of the MgO substrate edges is about 2.6% of that of the substrate surface area. The evaluated total emissivity of the edges is about 60% as calculated from the optical thickness of the MgO substrate when seen from the side and considering the influence of the Si rods that can be seen from the side. As a result, the power that is emitted from the edges at 1338 K becomes about 12.5% of $P_{\text{in}}^{\text{tot}}$. The calculated loss that is caused by the thermal radiation and heat conduction of the thin Pt wire at 1338 K is about 3.3% of $P_{\text{in}}^{\text{tot}}$. The energy loss that originates from the air molecules in the vicinity is only ≈0.04% owing to the measurement under a sufficiently low pressure ($3 \times 10^{-2}$ Pa), and thus can be omitted in this discussion. Furthermore, we can derive 2.5% for the loss caused by the view factor. When we evaluate the sum of these losses (18.3% of the total input power), we find $P_{\text{leak}}^{\text{tot}} = 0.183 \times 1.24$ W $= 0.227$ W for an emitter with $A_{\text{emitter}} = 0.230$ cm$^2$. If we assume that the power density of the above mentioned losses ($P_{\text{leak}} = P_{\text{leak}}^{\text{tot}}/$ $A_\text{emitter}$) can be reduced by upscaling, the heat flux density required to maintain an emitter temperature of 1338 K becomes 1.01 W/ 0.230 cm$^2$ =4.39 W/cm$^2$. When we set $A_\text{PV cell} = A_\text{emitter}$, the $P_\text{out}^\text{footptint}$ would be 0.139 W/ 0.230 cm$^2$ = 0.604 W/cm$^2$. However, we have to consider the effect of the increase in the open-circuit voltage upon enhancement of the short-circuit current density. The $P_\text{out}^\text{footptint}$ when considering this effect would be 0.638 W/cm$^2$. Therefore, the system efficiency can increase to 14.5% by upscaling.

Finally, we examine the influence of the aperture ratio, which is determined by the geometry of the contact grid of the PV cell. The present cell's aperture ratio is 77%, but it is possible to reach 100% by employing a back surface contact scheme (*28*). In case of no shadowing by the electrodes and considering the above mentioned effect, the output power increases by 33.5%. Hence, the $P_\text{out}^\text{footptint}$ that can be provided at 1338 K increases to 0.852 W/cm$^2$. Because the input heat flux density does not differ from the above derived 4.39 W/cm$^2$, the ideal system efficiency for the present PC-emitter design becomes 19.4%.

**Conclusion**

In conclusion, we designed an efficient thermophotovoltaic system that employs a Si rod-type PC thermal emitter, which has a relatively narrow emission spectrum. The employed emitter exhibited a high spectral efficiency of 42.6% with respect to the band-gap energy of the absorber layer ($\lambda_{\text{Eg}}$ = 1.76 μm). In order to quantitatively evaluate the input heat flux $P_{\text{in}}^{\text{tot}}$ and the not usable dissipated power $P_{\text{leak}}^{\text{tot}}$, we thermally isolated the emitter by suspending it in vacuum with thin Pt wires. We adopted the method of converting the input electrical power into a heat flux via an internal heater in order to have a well-defined heat source. For an emitter at 1338 K, we measured a high value of 11.2% for the actual system efficiency. This is 1.65 times the highest value among the values reported so far. The $P_{\text{out}}^{\text{footptint}}$ is measured to be 0.368 W/cm². The system efficiency estimated by removing losses that can be reduced by a scale up of the system size, is 14.5%. Furthermore, we estimated that a system efficiency of 19.4% can be reached if the contact grid design of the PV cell is improved. In this case, a high $P_{\text{out}}^{\text{footptint}}$ of 0.852 W/cm². is expected. Our experimental result can be considered as the present world record for the efficiency of a thermophotovoltaic system in which the input power has been directly measured. Furthermore, the predicted upscaled system efficiency exceeds the previously reported semi-empirical limit. We expect that by further improvements in the future, the heat that is usually considered wasted can be utilized more effectively. These improvements may aid the development of distributed energy supplies using combustion heat.

**Acknowledgments:** We thank K. Ishizaki for fruitful discussions. **Funding:** This work was supported in part by Japan Society for the Promotion of Science Grant-in-Aid for Scientific Research (KAKENHI) grant no. 17H06125. **Author contributions:** M.S. performed the system designs, analyses and experiments. T.A. and S.N. designed the study. T.I. examined the photovoltaic cell. All authors participated in discussing the results. **Competing interests:** The authors declare that they have no competing interests. **Data and materials availability:** All data needed to evaluate the conclusions in the paper are present in the paper s. Additional data related to this paper may be requested from the authors.

# Supplementary Materials for

**High-efficiency thermophotovoltaic system that employs an emitter based on a silicon rod-type photonic crystal**

Masahiro Suemitsu[1,2], Takashi Asano[1*], Takuya Inoue[1], Susumu Noda[1]

**Affiliations:**

[1]Department of Electronic Science and Engineering, Kyoto University, Kyoto 606-8501, Japan

[2]Energy Technology Laboratories, Osaka Gas Co., Ltd., 6-19-9 Torishima, Konohana-Ku, Osaka 554-0051, Japan

[*]Correspondence to: tasano@qoe.kuee.kyoto-u.ac.jp

**This PDF file includes:**

    Materials and Methods

**Materials and Methods**

**Fabrication of the thermal emitter**

We prepared a 20-nm thick $HfO_2$ layer on a 500-μm thick MgO substrate and then grew 825 nm of poly-crystalline Si using the LP-CVD method. The PC pattern was written by electron-beam lithography, and the Si rod structure was obtained by dry etching. The locations of the heater, the thermocouple, and their electrical contact pads were first patterned by photolithography, and the Si rod structure in these locations was removed by dry etching. After the resist mask had been removed, we fabricated a 30-nm thick $Al_2O_3$ layer on the whole emitter by atomic layer deposition to improve the heat resistance. For the formation of the heater and thermocouple elements, a film was fabricated by successively sputtering $HfO_2$ and $Al_2O_3$ (each time 50 nm). Then the patterns of the electrodes were written by photolithography, and finally the Pt/Ti layers with thicknesses of 300 and 30 nm were made by sputtering. The thermal emitter with the fabricated heater and thermocouple elements was cut into a square with size 4.8×4.8 mm, and thinned to 65 μm using a polishing machine. After polishing, Pt and $Pt_{0.9}Rh_{0.1}$ wires with $\phi$ 25 μm were attached to their corresponding contact pads by using a wedge bonder. We note that at the contact pads for the heater, we attached two Pt wires on each pad. With this four-terminal configuration, we accurately measured the voltage difference between the contact pads of the heater, an evaluated the electrical power consumed by the heater inside the thermal emitter. The wires attached to the contact pads are connected to a universal electrical board via soldering, and we adjusted each wire's length to 25 mm.

**Fabrication of the photovoltaic cell**

The PV cell used in this work is based on InGaAs. First, an n-type InP layer with thickness 0.1 μm and a doping concentration of n = $3\times10^{18}$ $cm^{-3}$ was epitaxially grown on a semi-insulating Fe-InP substrate by MOCVD. This was followed by the growth of n-type $In_{0.53}Ga_{0.47}As$ (n = $3\times10^{18}$ $cm^{-3}$, 0.3 μm), p-type $In_{0.53}Ga_{0.47}As$ (p = $2\times10^{17}$ $cm^{-3}$, 2.0 μm), p-type InP (p = $2\times10^{18}$ $cm^{-3}$, 0.1 μm), and finally a $p^+$-$In_{0.53}Ga_{0.47}As$ layer (p = $5\times10^{18}$ $cm^{-3}$, 0.2 μm). The area of the electrode for the p-side was defined by photolithography, and a Au/Pt/Ti layer structure was deposited by evaporation. The $p^+$-$In_{0.53}Ga_{0.47}As$ layer in the area outside the region of the electrode for the p-side was removed by etching. The area of the electrode for the n-side was patterned by photolithography, and the p-InP/ p-$In_{0.53}Ga_{0.47}As$/ n-$In_{0.53}Ga_{0.47}As$ structure was removed by etching. A Au/Pt/Ti layer structure was fabricated on the exposed n-InP area, and then the device was annealed at 673 K for 20 s using a rapid thermal processing (RTA) process. After annealing, the PV cell area without electrodes was coated with a 206-nm-thick $SiO_2$ layer in order to reduce the reflection of light with wavelengths shorter than 1.76 μm. After this, we increased the thickness of the electrode part to 7 μm by plating, which reduces the series resistance. The backside of the PV cell was prepared by depositing 150 nm Al via evaporation. The fabricated PV cell was attached to a gold-plated copper substrate using a conductive Ag paste (Fujikura Kasei, XA-874), and connected to an external source meter via 10 μm thick and 100 μm broad gold ribbons.